\documentclass[11pt,english]{article}
\usepackage[english]{babel}
\usepackage[T1]{fontenc}
\usepackage{authblk}
\usepackage[utf8]{inputenc}
\usepackage{newtxmath,newtxtext}
\usepackage[dvipsnames]{xcolor}
\usepackage{tabularx}

\usepackage{graphicx}
\usepackage{epsfig}
\usepackage[authoryear]{natbib}

 \bibpunct[, ]{(}{)}{,}{a}{}{,}%
 \def\BIBand{and}%

\usepackage{geometry}
\makeatletter

\usepackage{amsthm}
\usepackage{amsmath}
\usepackage{amsthm}
\usepackage{subfig}
\usepackage{mathtools}
\usepackage{bigints}
\usepackage{graphicx}
\usepackage{booktabs}
\usepackage[colorinlistoftodos]{todonotes}
\usepackage[colorlinks=true, allcolors=blue]{hyperref}
\usepackage{verbatim}
\usepackage{float}

\usepackage{amssymb}
\usepackage{graphicx}
\usepackage{setspace}
\usepackage[final]{pdfpages}
\usepackage{esint}
\newtheorem{corollary}{Corollary}
\newtheorem{proposition}{Proposition}
\usepackage{array,longtable}
\usepackage{titlesec}
\usepackage{ragged2e}
\usepackage{pgfplots}
\pgfplotsset{width=10cm,compat=1.9}
\usepgfplotslibrary{external}
\@ifundefined{showcaptionsetup}{}{%
 \PassOptionsToPackage{caption=false}{subfig}}
\makeatother

\begin{document}

\title{When Is Self-Disclosure Optimal? Incentives and Governance of AI-Generated Content}

\author{
Juan Wu$^{1}$, Zhe (James) Zhang $^{2}$, Amit Mehra$^{2}$
\\
{\small $^{1}$ University of Science and Technology of China \quad $^{2}$University of Texas at Dallas}
}

\date{}
\maketitle

\begin{abstract}
Generative artificial intelligence (Gen-AI) is reshaping content creation on digital platforms by reducing production costs and enabling scalable output of varying quality. In response, platforms have begun adopting disclosure policies that require creators to label AI-generated content, often supported by imperfect detection and penalties for non-compliance. This paper develops a formal model to study the economic implications of such disclosure regimes. We compare a non-disclosure benchmark, in which the platform alone detects AI usage, with a mandatory self-disclosure regime in which creators strategically choose whether to disclose or conceal AI use under imperfect enforcement. The model incorporates heterogeneous creators, viewer discounting of AI-labeled content, trust penalties following detected non-disclosure, and endogenous enforcement. The analysis shows that disclosure is optimal only when both the value of AI-generated content and its cost-saving advantage are intermediate. As AI capability improves, the platform’s optimal enforcement strategy evolves from strict deterrence to partial screening and eventual deregulation. While disclosure reliably increases transparency, it reduces aggregate creator surplus and can suppress high-quality AI content when AI is technologically advanced. Overall, the results characterize disclosure as a strategic governance instrument whose effectiveness depends on technological maturity and trust frictions.
\end{abstract}

\section{Introduction}

Generative artificial intelligence (Gen-AI) is rapidly transforming creative production on digital content platforms. Recent advances in large-scale generative models, such as GPT-5.2 and Gemini Ultra, have substantially expanded the scope, speed, and scalability of content creation. These technologies enable creators to produce text, images, audio, and video at unprecedented scale while significantly reducing production costs. For platforms, the diffusion of Gen-AI promises a larger and more diverse content supply, lower entry barriers for creators, and expanded monetization opportunities through increased engagement and advertising revenue.

At the same time, the proliferation of Gen-AI poses a fundamental governance challenge for content platforms. While AI tools can enhance productivity and content quality, they also raise concerns about authenticity, credibility, and user trust. Prior research shows that audiences often perceive human-generated content as more authentic and emotionally resonant, leading to higher engagement and attachment \citep{wan2017attachment,kim2019says}. As AI-generated content becomes increasingly human-like, however, distinguishing between human and machine authorship has become both technically difficult and strategically consequential. Viewers may value transparency regarding content creation, yet such transparency can also trigger skepticism and reduce engagement, particularly in credibility-sensitive domains.

Historically, platforms have relied on algorithmic detection to identify AI-generated or manipulated content, using linguistic patterns, metadata, and behavioral signals. However, detection accuracy remains imperfect and costly to maintain. Empirical evidence suggests that state-of-the-art detectors often achieve accuracy below 80\% and perform unevenly across languages, formats, and adversarial paraphrasing \citep{weber2023testing,sadasivan2023can,krishna2023paraphrasing}. Detection models also require continuous retraining and labeled data, making them an incomplete solution at scale or leading to huge costs. Moreover, detection alone does not address the strategic incentives of creators, who may choose whether to disclose or conceal AI usage based on expected audience responses and platform enforcement.

In response, platforms have increasingly turned to disclosure policies that require creators to self-report AI usage. Since early 2024, Meta has mandated labels such as ``Imagined with AI'' on Facebook and Instagram, while YouTube has implemented an ``Altered or Synthetic Content'' disclosure requirement for AI-generated or manipulated media.\footnote{\url{https://about.fb.com/news/2024/02/labeling-ai-generated-images-on-facebook-instagram-and-threads/}; \url{https://blog.youtube/news-and-events/disclosing-ai-generated-content/}} These policies typically combine voluntary self-disclosure with algorithmic detection as a backstop and impose penalties—such as reduced visibility, downranking, or removal—when undisclosed AI usage is detected.\footnote{ \url{https://about.fb.com/news/2024/04/metas-approach-to-labeling-ai-generated-content-and-manipulated-media/} and \url{https://transparency.meta.com/policies/community-standards/misinformation}.} Yet disclosure practices vary substantially across platforms, and several major platforms (e.g., X, Reddit, Bluesky) have not adopted mandatory disclosure regimes.

Despite their growing prevalence, the economic consequences of disclosure policies for AI-generated content remain poorly understood. Disclosure is typically framed as an ethical or transparency-enhancing intervention, yet it also operates as a strategic governance instrument that reshapes creator incentives, content composition, and platform value. By altering how AI adoption maps into engagement and revenue, disclosure affects not only whether creators use AI, but also how and when they choose to reveal that usage. Identical AI-generated content can generate markedly different engagement outcomes depending on whether AI use is voluntarily disclosed, revealed through platform detection, or strategically concealed. As a result, disclosure does not merely shift adoption margins; it fundamentally changes creators’ strategic calculus regarding transparency, concealment, and content positioning. In this sense, disclosure functions as an endogenous incentive mechanism that reallocates AI usage across creator types and content domains, rather than as a simple binary constraint on technology adoption.

This paper addresses a set of interrelated research questions concerning the governance of AI-generated content on digital platforms. 

1. \textit{Under what conditions should a platform mandate disclosure of AI usage rather than rely on algorithmic detection alone?} 

2. \textit{How do disclosure policies reshape creators’ strategic incentives, not only regarding whether to adopt Gen-AI, but also whether to disclose, conceal, or selectively reveal its use?}

3. \textit{How should platforms optimally design enforcement intensity, such as penalties for non-disclosure, as the quality and cost efficiency of Gen-AI technologies evolve over time?}

4. \textit{How do disclosure regimes affect platform profitability, creator surplus, and aggregate content quality, and when do transparency gains come at the cost of reduced ecosystem value?}

By answering these questions within a unified formal framework, we clarify when disclosure policies function as effective governance tools and when they instead become frictions that distort AI adoption and value creation. Specifically, we develop a formal model comparing two regimes: a non-disclosure benchmark, in which the platform alone detects and labels AI-generated content, and a self-disclosure regime, in which creators are required to label AI usage subject to imperfect detection and penalties for non-compliance. The model captures three key features of Gen-AI ecosystems. First, AI reduces production costs and may improve content quality, but the magnitude of these effects varies across creators. Second, viewers discount content labeled as AI-generated and further penalize creators whose non-disclosure is revealed by the platform, reflecting both algorithm aversion and trust concerns \citep{castelo2019task,logg2019algorithm,longoni2019resistance}. Third, detection is imperfect, creating strategic incentives for concealment.

The analysis delivers several insights. Disclosure is optimal only within an intermediate region in which AI-generated content is valuable but not dominant and cost savings are substantial but not overwhelming. In this region, disclosure improves transparency and reallocates AI usage across creator types in a way that preserves trust without severely distorting productive efficiency. Outside this region, disclosure becomes counterproductive. When AI quality is low, disclosure primarily suppresses adoption without improving content value. When AI quality is high, disclosure-induced credibility discounts and enforcement costs distort efficient AI adoption and reduce aggregate content quality.

The results further show that disclosure is not neutral across market participants. Although self-disclosure reliably increases transparency and can raise platform profit by mitigating trust externalities, it does so by shifting compliance and credibility costs onto creators. Aggregate creator surplus therefore declines under disclosure, even in regions where the policy is privately optimal for the platform. This divergence highlights disclosure as a strategic governance tool rather than a neutral transparency mechanism and raises questions about the long-run sustainability of disclosure-based governance in AI-intensive content markets.

This paper contributes to the literature on platform governance and AI-enabled production in three ways. First, it models disclosure as an endogenous incentive mechanism that shapes creators’ adoption, disclosure, and concealment decisions, rather than as a passive transparency requirement. Second, it integrates imperfect detection, viewer discounting, and heterogeneous creator credibility to capture how trust frictions interact with technological efficiency in AI content markets. Third, by jointly analyzing platform policy design and creator behavior, the paper provides a unified economic perspective on disclosure as a mechanism that reallocates AI usage across creators and content domains, extending existing models of information disclosure and platform regulation to the Gen-AI context.

The remainder of the paper is organized as follows. Section~2 reviews the related literature. Section~3 presents the model. Section~4 analyzes equilibrium outcomes under disclosure and non-disclosure regimes and characterizes the platform's optimal policy. Section~5 discusses managerial and policy implications and concludes.

\section{Literature review}

\subsection{Platform Responses to AI-Generated Content}

In response to the growing presence of AI-generated and manipulated content, digital platforms have adopted a variety of governance strategies, ranging from economic optimization to algorithmic enforcement. From an economic and behavioral perspective, \citet{papanastasiou2020fake} develop a social learning framework to examine how platforms dynamically adjust detection mechanisms based on environmental volatility and incentive structures, offering insights into strategic moderation under uncertainty. Extending this, \citet{liu2022implications} investigates how content moderation decisions vary under different monetization models, revealing that for advertising-based platforms, increased moderation intensity does not necessarily translate into optimal platform outcomes. Building on this framework, \citet{jimenez2023economics} conducts two field experiments on Twitter and finds that hate speech moderation has a suppressive effect on both online discourse and offline hate crimes, highlighting a trade-off between societal benefit and platform engagement. This economic tension is further explored by \citep{madio2024content}, who examine how platforms balance advertiser preferences for safe environments with user tolerance for intrusive advertising. Their analysis suggests that while stricter content moderation may enhance advertiser satisfaction, it may also inadvertently increase ad exposure for users, thereby diminishing user experience.

Complementing this strategic lens, a range of technical solutions has been proposed to detect and mitigate manipulated or synthetic content. \citet{cavusoglu2009configuration} examine the interaction between firewalls and intrusion detection systems (IDS), illustrating that these systems can serve as either complements or substitutes depending on configuration. Focusing on disinformation, \citet{perez2017automatic} build machine learning models capable of achieving 76\% accuracy in fake news detection, contributing to the development of scalable classification tools. More recently, \citet{kirchenbauer2023watermark}introduce a watermarking framework for proprietary language models, revealing both the promise and limitations of embedding traceable signals in AI outputs. Similarly, \citet{tiwari2024ai} leverage semi-supervised GAN architectures to detect synthetic videos. Finally, \citet{sietsema2023study} explore user-facing disclosure strategies, such as nutritional label-style annotations for news authenticity. Their findings indicate that the effectiveness of these labels is not intrinsic but mediated by individual-level attributes such as partisanship, political ideology, and cognitive sophistication. Moreover, unlabeled misinformation does not benefit from a credibility boost, suggesting that labeling schemes may have bounded influence.

These studies reveal the multidimensional nature of platform responses to AI-generated content. They underscore the strategic trade-offs between moderation intensity, user engagement, advertiser value, and technical feasibility—highlighting the need for holistic governance frameworks that align platform interests with broader public welfare in the AI-driven content landscape.

\subsection{AI usage incentive}

The rapid advancement of Large Language Models (LLMs) is fundamentally transforming human–machine collaboration, particularly in creative and knowledge-intensive tasks. Existing literature has explored both the evolving capabilities of LLMs and their implications for user performance across various collaboration modes.

LLMs have demonstrated remarkable competence in addressing a wide range of complex problems. \citet{noy2023experimental} prove AI may not replace professionals, but rather serve as a tool to boost overall productivity, especially to help less-skilled groups perform better. \citet{bubeck2023sparks} highlight that GPT-4 can address novel and challenging problems across mathematics, coding, vision, medicine, law, psychology, and other domains without requiring special prompting. \citet{chen2024large} further extend this perspective by demonstrating these models can assist users with a wide range of open tasks (such as question answering, customer support, summarization) and creative tasks (including poetry composition, songwriting, and marketing content creation). Research by \citet{eloundou2023gptsgptsearlylook} indicates that approximately 15\% of all worker tasks in the United States could be completed significantly faster while maintaining the same quality level with LLM assistance. When incorporating software and tools built upon LLMs, this proportion potentially increases to between 47\% and 56\%.

Despite their impressive capabilities, LLMs still face limitations in creativity. Several studies have focused on enhancing the creative performance of LLMs. For instance, \citet{bunescu2019learning} propose a two-model architecture comprising an audience model for learning expectations connected to a composer model for learning to surprise, aimed at enhancing the creative capabilities of AI systems. \citet{ouyang2022training} emphasize LLMs' instruction-following capabilities, and proposing an avenue for aligning language models with user intent on a wide range of tasks by fine-tuning with human feedback to align with their users. \citet{bunescu2019learning} point out that due to their auto-regressive nature, LLMs may be restricted to generating expected outputs as they are trained to adhere to existing data distributions, lacking the element of "surprise." \citet{chen2024large} further emphasize that having LLMs lead the creative process may not be optimal. In situations where creativity constitutes a crucial component of output quality, the anchoring effect can become particularly salient, limiting the creativity of the final output. Relying solely on LLMs to generate content can constrain creativity, especially for expert users.

LLMs offer new possibilities for human-AI collaboration, with significant advantages in improving work efficiency and assisting non-expert users. However, caution is necessary when relying on LLMs for creative tasks, particularly for professionals. The optimal collaboration model likely involves using LLMs as supplementary tools rather than complete replacements for human creative processes. Future research should focus on optimizing human-AI collaboration modalities to maximize LLMs' advantages while addressing their creative limitations.

\subsection{User Behavior and Platform Dynamics}
Recent research on AI-generated content has illuminated the complex interplay between content source, user perception, and behavior on digital platforms. Early investigations into algorithmic decision-making reveal initial user skepticism and resistance \citep{dietvorst2015algorithm,castelo2019task,longoni2019resistance,liu2023algorithm}, while other studies suggest that users can appreciate algorithmic outputs under specific conditions, particularly when such outputs exceed their expectations \citep{logg2019algorithm}. However, the deployment of Gen-AI technologies has raised new concerns about bias and authenticity. \citet{fang2024bias} provide empirical evidence that content generated by leading large language models (LLMs), such as ChatGPT and LLaMA, exhibits substantial gender and racial bias, heightening public concerns about fairness and representation. 

These concerns extend to both viewer and creator behavior on digital platforms. On the viewer side, \citet{luo2019frontiers} conducted an experiment showing that users responded similarly to human and chatbot sales agents before disclosure, but once the chatbot identity was revealed, purchase rates significantly declined. This indicates that perceived authenticity plays a critical role in shaping user behavior. \citet{turel2023prejudiced} further suggest that users may subconsciously perceive AI systems as alien entities, leading to prejudicial responses. Supporting this,  \citet{bruns2024you} through a series of experimental studies, find that Gen-AI disclosure can trigger negative attitudinal and behavioral reactions toward brands, primarily through diminished perceptions of authenticity—particularly when Gen-AI fully replaces human labor in content creation. \citet{WangMeixian} based on an online experiment reveals that disclosure violates consumers' expectations and evokes negative feelings. In contrast, \citet{schanke2024digital} find that disclosure does not always reduce trust, particularly in contexts like voice cloning where the functional benefit may override disclosure effects. Similarly, \citet{yin2024ai} show that while AI-generated messages can increase users' sense of being heard, this effect is attenuated when users learn the message was machine-generated.

Viewers' behaviors are also shaped by platform design and content formats. \citet{wang2022seeing} find that users are more likely to report text-based fake news compared to video-based content, suggesting that content modality influences credibility assessments and reporting behavior. Platform monetization mechanisms also alter engagement: \citet{ye2022monetization} show that introducing a paid Q\&A feature boosts both content production and user interaction, reflecting how financial incentives and interaction models shape viewer participation.

From the creators' perspective, identity and incentives affect both content dissemination and strategic behavior. \citet{han2020importance} demonstrate that the interaction between creator attributes and content characteristics significantly improves content diffusion predictions. \citet{chen2019monetary} find that monetary incentives increase the volume and diversity of content, but not necessarily its quality, particularly in financial advisory contexts. \citet{bhargava2022creator} highlights how platform-provided tools, such as ad management systems and revenue-sharing models, affect the concentration of market power among creators. \citet{ye2022monetization} additionally show that the benefits of Q\&A systems depend on the identity of the content responders, emphasizing the hierarchical nature of creator influence.

Collaboration modalities between humans and AI introduce another layer of complexity in user interaction. \citet{fugener2022cognitive} demonstrate that team performance improves when AI systems delegate tasks to humans, whereas reverse delegation yields no significant benefits. This finding emphasizes the importance of role positioning in human-AI collaboration. \citet{bruns2024you} find that positioning Gen-AI as an assistant rather than a replacement helps mitigate negative user perceptions. \citet{chen2024large} show that LLMs enhance user-generated content quality when acting as co-creators, but may reduce user satisfaction when used as ghostwriters. These impacts vary by user expertise; LLMs are especially beneficial in bridging skill gaps between novices and experts. Fortunately, work by \citet{yao2024human} suggests that a stable equilibrium between human- and AI-generated content is achievable. However, how platforms can attain such equilibrium across different Gen-AI application contexts remains an open question for future research. These studies reveal that user behavior toward Gen-AI is shaped by a combination of content source, disclosure, task delegation, and platform design. Perceived authenticity, identity signaling, and incentive mechanisms are central to understanding engagement, trust, and economic outcomes in AI-mediated content ecosystems.

\section{Model Setup}

\subsection{Platform, Creators, and Revenue Sharing}

We consider a content platform (e.g., YouTube or Facebook) that hosts a continuum of content creators. Creators publish digital content such as images, audio, or video, which the platform monetizes through advertising, subscriptions, donations, and related mechanisms. Total revenue generated by a piece of content is shared between the platform and the creator: the platform retains a fraction $r \in (0,1)$ as commission, while the creator receives the remaining fraction $1-r$. In practice, major platforms such as Meta and YouTube typically charge commission rates ranging from 30\% to 45\%. 

Moreover, each creator produces a single unit of content and earns revenue proportional to viewer engagement. We assume that viewer engagement is increasing in content quality, reflecting the fact that platform monetization mechanisms are driven by interaction intensity and attention. We therefore model revenue as a reduced-form increasing function of content quality, abstracting from the micro-level dynamics of engagement formation while preserving the core economic linkage between quality and monetization. Table~\ref{tab:notation} summarizes the main notation used in the model.

\begin{table}[ht]
\centering
\caption{Major Notation}
\label{tab:notation}
\renewcommand{\arraystretch}{0.9}
\setlength{\tabcolsep}{6pt}
\begin{tabularx}{\textwidth}{>{\centering\arraybackslash}p{2.5cm}X}
\toprule
\textbf{Notation} & \textbf{Definition} \\
\midrule
$v$ & Quality of AI-generated content \\
$c$ & Cost of producing human-generated content \\
$\delta$ & Cost reduction factor for AI-generated content, $\delta \in (0,1)$ \\
$\beta$ & Detection accuracy of the platform's algorithm \\
$r$ & Platform commission rate \\
$f$ & Viewer credibility discount for labeled AI-generated content \\
$k$ & Trust discount following detected non-disclosure \\
$p$ & Penalty for failing to disclose AI usage \\
$\Pi_i$ & Platform profit under policy $i \in \{N,D\}$ \\
\bottomrule
\end{tabularx}
\end{table}

\subsection{Content Creation Technologies}

Creators choose between two production technologies: human-generated content ($NAI$) and AI-generated content ($AI$).

Human-generated content incurs a production cost $c>0$ and delivers normalized quality equal to one. AI-generated content reduces production costs but may alter content quality. Specifically, using AI incurs a reduced cost $\delta c$, where $\delta \in (0,1)$ captures the extent to which GenAI substitutes for human effort.

We denote the quality of AI-generated content by $v$, where $v \in (0,\bar v)$ and $\bar v > 1$. This formulation captures the idea that GenAI may underperform or outperform human creation depending on the model used and the application context. Higher content quality translates into greater viewer engagement, which increases monetization opportunities on the platform. When $v>1$, AI-generated content generates higher gross engagement than human-generated content; when $v<1$, human creation dominates in quality.

\subsection{Platform Moderation and Disclosure Policies}

The platform moderates content using an automated detection algorithm that attempts to identify AI-generated content. We consider two moderation regimes.

Under the \emph{non-disclosure regime} ($i=N$), the platform relies solely on its internal detection system to label AI-generated content. Creators are not required to disclose AI usage.

Under the \emph{self-disclosure regime} ($i=D$), creators are required to truthfully label AI-generated content. If a creator fails to disclose and the platform subsequently detects AI usage, the platform imposes a penalty $p \ge 0$, which may represent removal, downranking, demonetization, or account-level sanctions.

The detection algorithm correctly classifies content with probability $\beta \in \left(\tfrac{1}{2},1\right)$. For simplicity, detection accuracy is assumed to be symmetric: AI-generated and human-generated content are each correctly identified with probability $\beta$. This assumption allows us to abstract away from classification bias and focus on the platform's mechanism design problem induced by disclosure requirements. We hold $\beta$ fixed across regimes to isolate the incentive effects of disclosure policy rather than technological differences in detection performance.

To abstract from intrinsic discrimination against AI content, we normalize any direct penalty imposed on detected AI content under the non-disclosure regime to zero.

\subsection{Viewer Perceptions and AI Aversion}

Viewers derive utility from consuming content but may discount content that is explicitly labeled as AI-generated. This discount reflects concerns about authenticity, bias, and accountability, and varies across viewers and content domains.

We capture this heterogeneity through a credibility discount factor $f \in [0,1]$, drawn from a uniform distribution. This specification parsimoniously captures heterogeneity in viewer receptivity to AI-generated content while ensuring analytical transparency. Our results rely on the presence of heterogeneity rather than on the specific functional form of the distribution. When content is labeled as AI-generated, viewers value it at a fraction $f$ of its underlying quality $v$. A lower value of $f$ corresponds to stronger aversion to AI-generated content. This perceptual discount applies whenever AI usage is revealed, regardless of whether disclosure is voluntary or system-detected.

\subsection{Trustworthiness and Disclosure Compliance}

Disclosure policies may also affect perceptions of creator trustworthiness. When a creator fails to disclose AI usage and is subsequently flagged by the platform, viewers may infer intentional concealment, leading to reputational harm.

We capture this effect through a trust discount factor $k \in (0,1)$. When AI usage is detected following non-disclosure, viewers discount content value by an additional multiplicative factor $k$. This trust penalty does not apply when AI usage is truthfully disclosed.

\subsection{Creator Payoffs}

Creators maximize expected utility from content publication, which depends on revenue net of platform commission and production costs.

\paragraph{Non-disclosure regime.}
Under the non-disclosure regime, creators choose between human-generated and AI-generated content. Expected utilities are given by
\begin{equation}
\label{utilityN}
\begin{aligned}
u_{NAI} &= 1 - r - c, \\
u_{AI} &= (1-r)\big[\beta f v + (1-\beta)v\big] - \delta c .
\end{aligned}
\end{equation}

When AI-generated content is detected, it is discounted according to $f$; when it is not detected, it is evaluated solely based on quality $v$. Human-generated content is assumed to be perfectly identified after appeal, which we normalize to be costless.

\paragraph{Self-disclosure regime.}
Under mandatory disclosure, creators have three strategies: use no AI ($NAI$), use AI without disclosure ($AINAI$), or use AI with truthful disclosure ($AIAI$). Expected utilities are
\begin{equation}
\label{utilityD}
\begin{aligned}
u_{NAI} &= 1 - r - c, \\
u_{AINAI} &= (1-r)\big[\beta(fkv - p) + (1-\beta)v\big] - \delta c, \\
u_{AIAI} &= (1-r) f v - \delta c .
\end{aligned}
\end{equation}

Relative to the non-disclosure regime, noncompliance under disclosure exposes creators to both reputational loss ($k$) and penalties ($p$) when detected. Truthful disclosure avoids these losses but subjects creators to viewer discrimination against AI-generated content with certainty.

We restrict attention to the non-trivial parameter region in which AI-generated content is economically viable absent disclosure frictions but does not dominate human creation regardless of policy, ensuring that disclosure incentives play a meaningful role. Specifically, we assume
\[
\underline v_1 \equiv \frac{1-r-c(1-\delta)}{1-r} < v.
\]

If $v<\underline v_1$, creators never adopt AI and disclosure policy is irrelevant: all equilibrium results collapse to the human-made content only case.

We treat the disclosure regime as a policy chosen by the platform prior to content creation. To clarify economic mechanisms, we first solve the equilibrium under each regime separately, and then compare platform payoffs to characterize the optimal policy choice.

\subsection{Timeline}

The sequence of events unfolds as follows:\\
1. \textbf{Platform decision.} The platform chooses whether to implement a self-disclosure policy. If so, it sets the penalty $p$ for noncompliance.\\
2. \textbf{Creator decision.} Creators choose their production technology and, under disclosure, whether to truthfully disclose AI usage.\\
3. \textbf{Viewer response.} Viewers observe content and labels, form perceptions about content quality and creator trustworthiness, and decide their level of engagement.

\section{Equilibrium Analysis}

We characterize the subgame perfect equilibrium by backward induction. In Stage~3, viewers respond to the observed content and label, valuing AI-labeled content according to the credibility discount $f$ and, when applicable, the trust discount $k$. In Stage~2, creators choose a creation technology and, under mandatory disclosure, whether to truthfully disclose AI usage. In Stage~1, the platform selects the moderation regime and, under disclosure, the penalty $p$ for detected noncompliance.

\subsection{Benchmark: Non-Disclosure Regime}

Under non-disclosure, creators choose between $NAI$ and $AI$ by comparing utilities in Equation \eqref{utilityN}. Because AI content is discounted only when labeled, creator incentives depend on the receptivity draw $f$: creators in authenticity-sensitive domains (low $f$) are less willing to adopt AI, whereas creators in domains with higher acceptance of AI (high $f$) are more willing to adopt.

\begin{figure}[ht] \centering \includegraphics[width=0.6\textwidth]{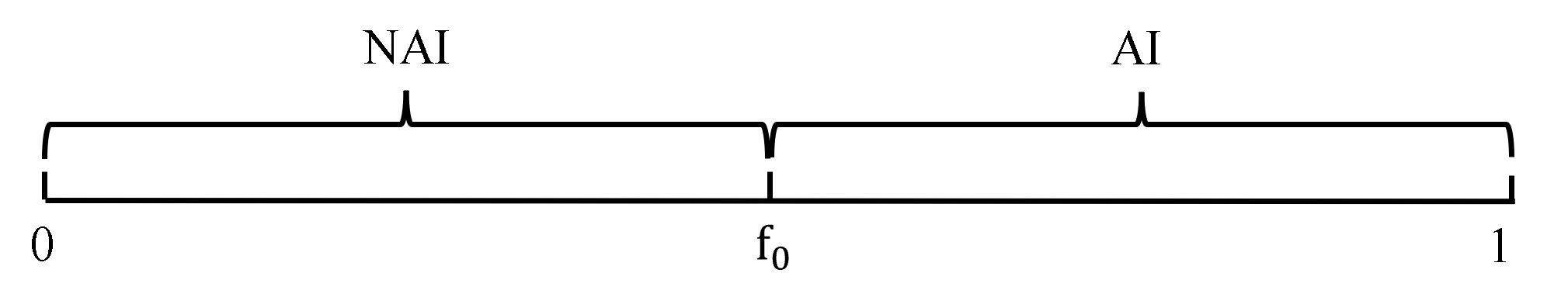} \caption{\centering Creators' choice of using Gen-AI under non-disclosure policy} \label{fig:1}
\end{figure}

\paragraph{Creator cutoff and segmentation.}
When AI quality is intermediate, $\underline{v}_1<v<\hat v$, equilibrium exhibits a cutoff $f_0\in(0,1)$ such that creators with $f<f_0$ choose $NAI$ and creators with $f\ge f_0$ choose $AI$. This sorting is depicted in Figure~\ref{fig:1}. Formally,
\[
f_0=\frac{(1-r)\big(1-v(1-\beta)\big)-c(1-\delta)}{(1-r)v\beta},
\qquad 
\hat v=\frac{1-r-c(1-\delta)}{(1-r)(1-\beta)}.
\]
When AI quality is sufficiently high, $\hat v<v<\bar v$, the benefits from higher $v$ and lower cost $\delta c$ dominate the expected discount from being labeled, and all creators adopt AI. In Figure~\ref{fig:1}, the cutoff collapses to $f_0=0$.

We use subscript $N$ to denote the non-disclosure regime and use $N_N(i)$ to denote the measure of creators choosing $i\in\{NAI,AI\}$. Then for $\underline{v}_1<v<\hat v$,
\[
N_N(NAI)=f_0,\qquad N_N(AI)=1-f_0,
\]
and for $\hat v<v<\bar v$, $N_N(AI)=1$.

\paragraph{Platform profit and comparative statics.}
Aggregating commission revenue over creators yields:

\begin{equation}
\label{eqa:PiN}
\Pi_N=
\begin{cases}
\displaystyle 
\frac{r \left((1-r)^2 \left(2 \beta  v+(1-v)^2\right)-c^2 (1-\delta)^2\right)}{2 \beta  (1-r)^2 v}, 
& \underline{v}_1<v<\hat{v},\\[8pt]
\displaystyle 
\frac{(2-\beta) r v}{2}, 
& \hat{v}<v<\overline{v}.\\
\end{cases}
\end{equation}

Proposition~\ref{prop:1} then characterizes how platform profit varies with AI value $v$ and detection accuracy $\beta$.

\begin{proposition}
\label{prop:1}
Under the non-disclosure regime\\
(i) \label{prop:1i} platform profit decreases in $v$ for $\underline{v}_1<v<\min\{\tilde{v}_1,\hat{v}\}$ and increases in $v$ for $\min\{\tilde{v}_1,\hat{v}\}<v<\bar{v}$.\\
(ii) \label{prop:1ii} platform profit increases in $\beta$ for $\underline{v}_1<v<\min\{\tilde{v}_2,\hat{v}\}$ and decreases in $\beta$ for $\min\{\tilde{v}_2,\hat{v}\}<v<\bar{v}$.

The thresholds $\tilde{v}_1$ and $\tilde{v}_2$ are provided in the Appendix.
\end{proposition}

The economic intuition is consistent with the sorting in Figure~\ref{fig:1}. When $v$ is in an intermediate range, a marginal increase in $v$ attracts additional creators into AI adoption (a smaller $f_0$), but the marginal AI content that replaces human creation is still discounted with high probability, reducing realized engagement. Since the platform does not internalize creators' cost savings, platform profit can fall even as creators become more eager to adopt AI. Once $v$ is sufficiently large, AI content remains valuable even when labeled and discounted, and higher $v$ increases platform profit.

The effect of detection accuracy $\beta$ is likewise non-monotone. When AI content is not very valuable, higher $\beta$ discourages adoption among low-$f$ creators and improves the average engagement of the content pool, raising profit. When AI content is valuable, higher $\beta$ increases the incidence of labeling and discounting and reduces the realized value of AI content, lowering profit.

\subsection{Mandatory Disclosure Regime}
\label{subsec:disclosure}

We now analyze the equilibrium when the platform mandates disclosure of AI usage. Under this regime, creators choose among $NAI$ (no AI), $AINAI$ (use AI and conceal), and $AIAI$ (use AI and disclose), as characterized by the utilities in~\eqref{utilityD}. Creators' choices depend on AI content value $v$ and the penalty $p$ imposed when AI usage is detected following non-disclosure.

\begin{figure}[ht]
  \centering
  \includegraphics[width=0.6\textwidth]{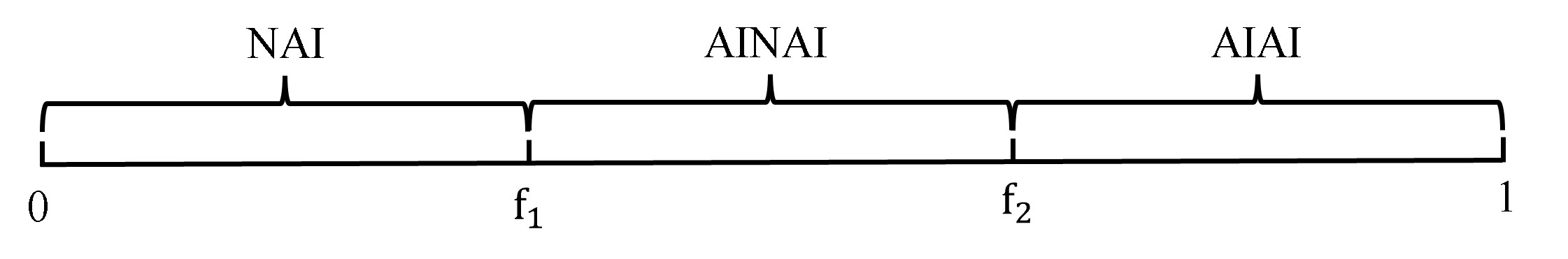}
  \caption{\centering the creators' segmentation under disclosure case}
  \label{fig:2}
\end{figure}

\paragraph{Creator segmentation in $f$ (Figure~\ref{fig:2}).}
For a given $(v,p)$, equilibrium exhibits cutoff-based sorting in the credibility parameter $f\in[0,1]$. Figure~\ref{fig:2} illustrates the canonical three-segment structure when all strategies are present: low-$f$ creators choose  $NAI$, intermediate-$f$ creators conceal ($AINAI$), and high-$f$ creators disclose ($AIAI$). The relevant cutoff points are
\[
f_1=\frac{(1-r)\big(1-v(1-\beta)+p\beta\big)-c (1-\delta)}{(1-r)k v\beta},\qquad
f_2=\frac{v-(p+v)\beta}{v-k v\beta},\qquad
f_{12}=\frac{1-r-c (1-\delta)}{(1-r) v},
\]
where $f_1$ is the indifference point between $NAI$ and $AINAI$, $f_2$ is the indifference point between $AINAI$ and $AIAI$, and $f_{12}$ is the indifference point between $NAI$ and $AIAI$.

We also define
\[
\underline{v}_2=\frac{(1-k \beta)\big(1-r-c (1-\delta)\big)}{(1-\beta) (1-r)},\qquad
\hat{p}=\frac{c (1-\delta )+\big((1-\beta ) v-1\big) (1-r)}{\beta  (1-r)},\]
\[\qquad
\overline{p}=\frac{c (1-\delta) (1-\beta  k)+(1-r) (\beta  k-\beta  v+v-1)}{\beta  r}.
\]

\paragraph{Platform profit conditional on $p$.}
Given equilibrium segmentation, platform profit equals commission revenue integrated over realized engagement. We use subscript $D$ to denote the self-disclosure regime. Depending on the induced strategy profile, platform profit can be written as follows.

\noindent\textbf{Case 1:} $\underline{v}_1<v\leq\underline{v}_2$,
\begin{equation}
\label{eqa:4}
\Pi_D(p)= r\left(\int_0^{f_{12}}1\,df+\int_{f_{12}}^{1} f v \,df\right).
\end{equation}

\noindent\textbf{Case 2:} $\underline{v}_2<v<\hat{v}$,
\begin{equation}
\label{eqa:5}
\Pi_D(p)= 
\begin{cases}
\displaystyle
r\left(\int_0^{f_1}1 \, df+\int_{f_1}^{f_2} \big(\beta f k v+(1-\beta ) v\big) \, df+\int_{f_2}^1 f v\, df\right), & 0\leq p<\overline{p},\\[6pt]
\displaystyle
r\left(\int_0^{f_{12}}1 \, df+\int_{f_{12}}^1 f v \, df\right), & p\geq \overline{p}.
\end{cases}
\end{equation}

\noindent\textbf{Case 3:} $v\geq\hat{v}$,
\begin{equation}
\label{eqa:6}
\Pi_D(p)= 
\begin{cases}
\displaystyle
r\left(\int_0^{f_{2}} \big(\beta f k v+(1-\beta ) v\big) \, df+\int_{f_{2}}^1 f v\, df\right), & 0\leq p\leq\hat{p},\\[6pt]
\displaystyle
r\left(\int_0^{f_1}1 \, df+\int_{f_1}^{f_2} \big(\beta f k v+(1-\beta ) v\big)\, df+\int_{f_2}^1 f v\, df\right), & \hat{p}<p<\overline{p},\\[6pt]
\displaystyle
r\left(\int_0^{f_{12}}1 \, df+\int_{f_{12}}^1 f v \, df\right), & p \geq \overline{p}.
\end{cases}
\end{equation}

The platform chooses $p$ to maximize $\Pi_D(p)$ in~\eqref{eqa:4}--\eqref{eqa:6}. The next proposition characterizes the platform's optimal penalty.

\begin{proposition}
\label{prop:2}
Under the mandatory disclosure regime, the platform's equilibrium optimal penalty is
\begin{equation}
\label{eqa:7}
p^*= 
\begin{cases}
\displaystyle
\frac{c (1-\delta ) (1-\beta  k)+(1-r) (\beta  k-1+(1-\beta)v)}{\beta  (1-r)}, 
& \underline{v}_2<v<\min\{\tilde{v}_3,\tilde{v}_4\},\\[10pt]
\displaystyle
\frac{c (1-\delta) (1-\beta  k)}{\beta  (1-r)},  
& \tilde{v}_3<v<\tilde{v}_5,\\[10pt]
0, 
& \tilde{v}_4<v<\tilde{v}_3 \ \ \text{or}\ \ \max\{\tilde{v}_3,\tilde{v}_5\}<v<\hat{v}.\\
\end{cases}
\end{equation}
\end{proposition}
where $\tilde{v}_3$, $\tilde{v}_4$, and $\tilde{v}_5$ are defined in the Appendix.

\begin{figure}[ht]
  \centering
  \includegraphics[width=0.6  \textwidth]{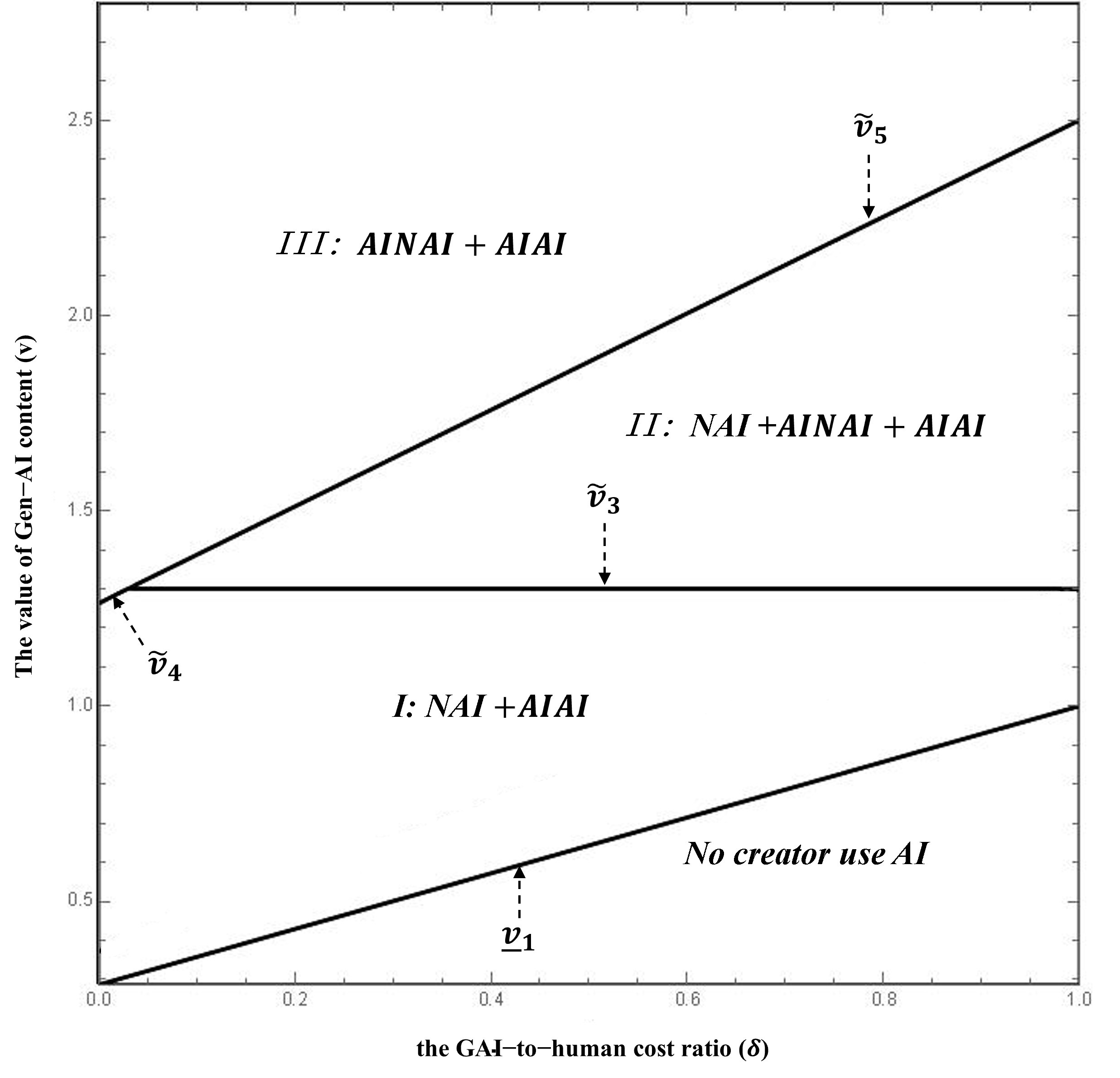}
  \caption{\centering Creators' equilibrium strategies under Mandatory Disclosure Regime when $k=0.8,\beta=0.6,r=0.3,c=0.5$}
  \label{fig:3}
\end{figure}

Proposition~\ref{prop:2} implies that optimal enforcement varies across the equilibrium regions depicted in Figure~\ref{fig:3}. When $v$ is relatively low (Region~I in Figure~\ref{fig:3}), equilibrium features only $NAI$ and $AIAI$: concealment does not arise (Figure~\ref{fig:2} collapses to two segments), and enforcement mainly serves to preserve transparency by preventing a nascent $AINAI$ segment from emerging. Accordingly, the platform sets a high penalty to deter concealment.

When $v$ is moderate (Region~II in Figure~\ref{fig:3}), concealment becomes privately attractive for an intermediate set of creators, yielding the three-way segmentation in Figure~\ref{fig:2}. In this region, fully eliminating $AINAI$ would require excessively high penalties that also suppress valuable AI adoption. The platform therefore chooses an intermediate penalty, trading off the screening of low-credibility creators against the retention of high-value AI content on the platform.

When $v$ is sufficiently high (Region~III in Figure~\ref{fig:3}), human creation is crowded out and equilibrium features only $AINAI$ and $AIAI$. In this region, penalties become counterproductive because they reduce the realized value of AI content by triggering label-induced discounts and trust losses upon detection. The platform optimally sets $p^*=0$ and allows creators to sort endogenously between concealment and disclosure.

\begin{corollary}
\label{cor:creator_strategies}
Under the mandatory disclosure regime, the equilibrium strategy set evolves across the regions in Figure~\ref{fig:3}:
\begin{enumerate}
\item If $\underline{v}_1<v<\tilde{v}_3$ (Region~I), creators choose $\{NAI, AIAI\}$.
\item If $\tilde{v}_3<v<\tilde{v}_5$ (Region~II), creators choose $\{NAI, AINAI, AIAI\}$.
\item If $v>\tilde{v}_5$ (Region~III), creators choose $\{AINAI, AIAI\}$.
\end{enumerate}
\end{corollary}

Corollary~\ref{cor:creator_strategies} formalizes the progression in Figure~\ref{fig:3}. As AI becomes more valuable (and/or more cost efficient as $\delta$ decreases), equilibrium transitions from a transparent adoption phase (Region~I), to a mixed phase with strategic concealment (Region~II), and eventually to an AI-dominant phase in which human creation disappears (Region~III). Figure~\ref{fig:2} provides the corresponding within-region sorting in $f$.

Substituting $p^*$ into~\eqref{eqa:4}--\eqref{eqa:6} yields the platform's equilibrium profit under mandatory disclosure:
\begin{equation}
\Pi_D^*= 
\begin{cases}
\label{eqa:PiD}
\frac{r \left((1-r)^2 \left(v^2+1\right)-c^2 (1-\delta)^2\right)}{2 (1-r)^2 v} &\text{if } \underline{v_1}<v<\min\{\tilde{v}_3,\tilde{v}_4\}\\
\frac{r ((1-r)^2 \left(v^2 \left(1+\beta  \left(\beta -\beta  k^2+k-2\right)\right)+(1-\beta  k)(1-2 (1-\beta) v)\right)-\beta  c^2 k (1-\delta)^2 (1-\beta  k))}{2 \beta  k (1-r)^2  (1-\beta  k)v}   & \text{if } \tilde{v}_3<v<\tilde{v}_5\\
\frac{r v (2-(2-\beta +k)\beta)}{2(1-\beta k)} & \text{if } \tilde{v}_4<v<\tilde{v}_3||\max\{\tilde{v}_3,\tilde{v_5}\}<v<\overline{v}.\\
\end{cases}
\end{equation}

After characterizing equilibrium enforcement, we next study how AI content value $v$ and detection accuracy $\beta$ affect $\Pi_D^*$.

\begin{proposition}
\label{prop:3}
Under the mandatory self-disclosure policy, the platform's equilibrium profit exhibits the following comparative statics:
\begin{enumerate}
\item \textbf{AI content value ($v$).}
If AI generates substantial cost savings ($0<\delta<\delta_0$), platform profit is decreasing in $v$ for $\underline{v}_1<v<v_D$ and increasing in $v$ for $v_D<v<\overline{v}$.  
If cost savings are limited ($\delta_0<\delta<1$), platform profit decreases in $v$ for $\underline{v}_1<v<v_D$, increases for $v_D<v<\tilde{v}_3$, decreases again for $\tilde{v}_3<v<\min\{v_8,\tilde{v}_5\}$, and finally increases for $\min\{v_8,\tilde{v}_5\}<v<\overline{v}$.

\item \textbf{Detection accuracy ($\beta$).}
Platform profit is independent of $\beta$ when $\underline{v}_1<v<\min\{\tilde{v}_3,\tilde{v}_4\}$, and strictly decreasing in $\beta$ when $\min\{\tilde{v}_3,\tilde{v}_4\}<v<\overline{v}$.
\end{enumerate}
\end{proposition}

Proposition~\ref{prop:3}(i) shows that platform profit evolves non-monotonically with the value of AI-generated content under mandatory disclosure. When AI quality is initially low, an increase in $v$ induces additional AI adoption by marginal creators whose content remains relatively low in realized value due to disclosure-induced credibility discounts and trust penalties. As a result, the average quality of content on the platform declines, leading platform profit to decrease despite technological improvement.

As $v$ continues to rise, the intrinsic quality of AI-generated content eventually dominates these credibility losses, and platform profit begins to increase. When AI cost savings are substantial ($\delta<\delta_0$), this transition generates a single turning point, yielding a U-shaped relationship between profit and $v$, mirroring the intuition in Proposition~\ref{prop:1}(i). In contrast, when AI cost savings are limited ($\delta>\delta_0$), the platform optimally employs an intermediate penalty over a range of $v$. This endogenous adjustment creates an additional decline in profit: as AI capability improves, a growing mass of intermediate-credibility creators adopts strategic non-disclosure ($AINAI$), diluting content quality before profit eventually recovers once penalties are fully relaxed. The resulting pattern is W-shaped.

Proposition~\ref{prop:3}(ii) characterizes the role of detection accuracy. When AI-generated content value is low, equilibrium creator strategies do not depend on detection precision, rendering platform profit invariant to $\beta$. Once AI content becomes sufficiently valuable, however, higher detection accuracy strictly reduces platform profit. Improved detection increases the likelihood that high-credibility creators are flagged when concealing AI usage, thereby subjecting their content to credibility and trust discounts and reducing its realized value. Consequently, stricter detection constrains the supply of high-quality AI content on the platform, lowering overall profitability.

\subsection{Platform's Optimal Regime Choice}

We next compare equilibrium profits across regimes and characterize the platform's optimal disclosure strategy. Proposition \ref{prop:4_restate} summarizes the equilibrium regime choice in the $(\delta,v)$ space, with the disclosure regime optimal only in an intermediate region.

\begin{figure}[ht]
  \centering
      \includegraphics[width=0.6\textwidth]{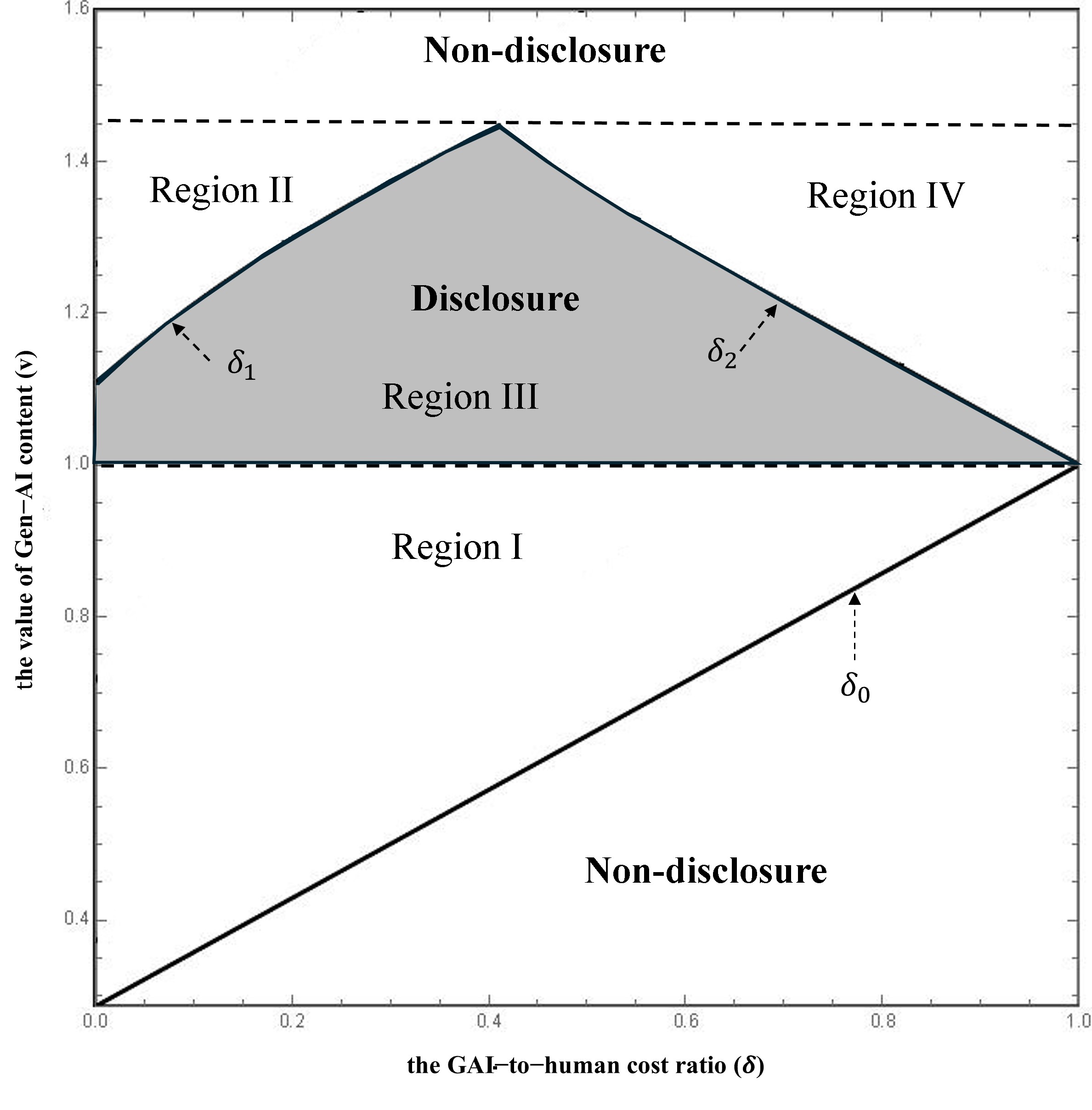}
  \caption{\centering Platform's optimal disclosure decision when $k=0.8,\beta=0.6,r=0.3,c=0.5$}
  \label{fig:4}
\end{figure}

\begin{proposition}[Platform's optimal disclosure strategy]
\label{prop:4_restate}
The platform adopts mandatory disclosure when AI content value and AI cost efficiency are both moderate (i.e., $\max\{\delta_1,0\}<\delta<\min\{\delta_0,\delta_2\}$). The platform prefers the non-disclosure regime, otherwise.
\end{proposition}

Figure~\ref{fig:4} provides the corresponding region interpretation. In \textbf{Region I} (low $v$), the platform prefers non-disclosure because AI content is not sufficiently valuable and disclosure primarily suppresses adoption without delivering commensurate gains in engagement. In \textbf{Region II} (very high $v$), the platform again prefers non-disclosure because disclosure and enforcement would unnecessarily restrict high-value AI adoption and intensify label-induced discount. In \textbf{Region III} (the shaded wedge), disclosure is optimal: it functions as a screening device that limits adoption among authenticity-sensitive creators and reduces the prevalence of low-realized-value AI content, improving aggregate engagement. In \textbf{Region IV} (high $\delta$, limited cost savings), non-disclosure dominates because disclosure further discourages adoption even though the technology provides relatively little cost advantage.

\subsection{Implications for Creators, Transparency, and Content Quality}

We next examine how the platform’s disclosure regime affects creators, informational transparency, and aggregate content quality. Creator profit is of particular interest because creators constitute the primary supply side of the platform. Their participation, effort, and long-run investment in content quality depend critically on expected returns. Disclosure policies that systematically alter creator surplus may therefore shape not only short-run outcomes but also the sustainability of the content ecosystem.

\paragraph{Creator profit.}
Let $CS_i$ denote aggregate creator profit under regime $i\in\{N,D\}$. Using equilibrium strategies,
\begin{equation}
\label{eqa:CS_restate}
\begin{aligned}
CS_N
&=\int_{0}^{f_0} \big((1-r)-c\big)\, df
+\int_{f_0}^{1} \Big((1-r)\big(\beta f v + (1-\beta)v\big) - \delta c\Big)\, df,\\
CS_D
&=\int_{0}^{f_1} \big((1-r)-c\big)\, df
+\int_{f_1}^{f_2} \Big((1-r)\big(\beta (fkv-p) + (1-\beta)v\big) - \delta c\Big)\, df
+\int_{f_2}^{1} \Big((1-r) f v - \delta c\Big)\, df.
\end{aligned}
\end{equation}

Based on Equations \ref{eqa:CS_restate}, we analyze the effects of self disclosure on the creator profit, as summarized in Proposition \ref{disclosureoncreator}.

\begin{proposition}
The platform's disclosure policy reduces the total creator profit. (i.e.,$CS_D<CS_N$).
\label{disclosureoncreator}
\end{proposition}

Although self-disclosure is commonly framed as a transparency-enhancing policy, our analysis, Propositions \ref{prop:4_restate} and \ref{disclosureoncreator}, shows that it also functions as a strategic governance instrument. By imposing disclosure-related costs on creators, the platform internalizes trust externalities at the ecosystem level, increasing its own payoff even if aggregate creator surplus declines.

\paragraph{Transparency and aggregate content quality.}
Following \citet{Zhu01012002}, we define transparency as one minus the probability that AI content appears without being labeled:
\begin{equation}
\label{eqa:T_restate}
\begin{aligned}
T_N &= 1-(1-\beta)\,N_N(AI),\\
T_D &= 1-(1-\beta)\,N_D(AINAI).
\end{aligned}
\end{equation}
Disclosure increases transparency ($T_D>T_N$) by reducing the mass of undisclosed AI content that escapes detection.

We next define aggregate content quality as the sum of quality contributions across all creators:

\begin{equation}
\label{eqa:Q_restate}
\begin{aligned}
Q_N &= N_N(NAI) + v\,N_N(AI),\\
Q_D &= N_D(NAI) + v\,N_D(AINAI) + v\,N_D(AIAI).
\end{aligned}
\end{equation}

\begin{proposition}
\label{qualitytransparency}
The platform’s disclosure policy affects transparency and aggregate content quality as follows:
\begin{enumerate}
    \item Disclosure strictly increases transparency (i.e., $T_D>T_N$).
    \item Disclosure improves aggregate content quality if and only if AI-generated content is technologically inferior to human creation (i.e., $Q_D>Q_N$ for $\underline{v}_1<v<\min\{\tilde{v}_4,1\}$), and reduces aggregate content quality when AI-generated content is technologically superior (i.e., $Q_D<Q_N$ for ($1<v<\tilde{v}_4||1<v<\tilde{v}_5)$). Otherwise, $Q_D=Q_N$.
\end{enumerate}
\end{proposition}

The quality result reflects a composition effect, which arises because disclosure alters the equilibrium allocation of production technologies across creators. By increasing the expected cost of AI usage, disclosure discourages AI adoption among marginal creators and shifts production toward human creation or labeled AI with lower realized engagement. Aggregate content quality therefore changes due to a reweighting of who produces content and how. When AI quality is low ($v<1$), this reallocation improves the content pool by crowding out inferior AI-generated content. When AI quality is high ($v>1$), the same mechanism suppresses superior AI output, lowering aggregate content quality despite higher transparency.

This results highlight that disclosure-based governance entails a trade-off between transparency and productive efficiency. While self-disclosure reliably reduces informational opacity by limiting undisclosed AI usage, it also raises the effective cost of AI adoption. When AI quality is low, this distortion is beneficial, as it crowds out inferior AI content and improves aggregate quality. When AI quality is high, the same distortion suppresses superior AI production, lowering content quality even as transparency improves. Only when AI quality and cost efficiency are intermediate does disclosure function as an effective screening device, enhancing transparency while preserving productive efficiency.

\section{Conclusion}
Based on the equilibrium analysis, we draw several conclusions regarding the governance of AI-generated content on digital platforms. The central insight is that disclosure and enforcement policies shape outcomes not only by improving information, but by reallocating production technologies and surplus across creators, with consequences that depend critically on the technological maturity of AI.

First, while human-generated content is eventually crowded out as AI quality improves, the path to this outcome depends on the cost efficiency of AI tools. When AI delivers substantial cost savings, the platform optimally abandons enforcement abruptly, removing penalties to fully exploit the scale and productivity gains from widespread AI adoption. When cost efficiency is more limited, the platform instead relaxes enforcement gradually. In this case, penalties serve as a temporary screening device that restrains low-credibility AI adoption while allowing incremental improvements in AI quality to be absorbed. This intermediate regime smooths the transition toward AI dominance by balancing short-run trust preservation against longer-run efficiency gains.

Second, the platform adopts a self-disclosure regime only when both the intrinsic value of AI-generated content and its cost-saving advantage lie in an intermediate range. In this region, disclosure operates as an effective governance instrument: it increases transparency, reallocates AI usage toward creators whose content is less sensitive to credibility concerns, and preserves aggregate content quality. Outside this range, disclosure becomes counterproductive. When AI quality is low, disclosure primarily suppresses adoption without delivering meaningful quality improvements. When AI quality is high, disclosure distorts efficient adoption by imposing valuation discounts on superior AI content, reducing aggregate content quality and creator surplus despite higher transparency.

Finally, our analysis shows that self-disclosure, though designed to enhance transparency, is not neutral across market participants. Platforms deploy disclosure strategically to manage trust externalities and content composition, often shifting compliance and credibility costs onto creators. As a result, disclosure can increase platform profit while reducing creator surplus and, in some regions, suppressing high-quality AI content. This tension suggests that transparency gains may come at the expense of creator incentives and aggregate content quality, raising concerns about the long-run sustainability of disclosure-driven governance.



\bibliographystyle{informs2014} 



%
%
%

\section*{Appendix: Proofs}

\subsection*{Proof of Proposition \ref{prop:1}}

When $0<v<\underline{v}_1$, $\frac{\partial \Pi_D}{\partial v}=0$, the profit is independent on $v$. When $\underline{v}_1<v<\hat{v}$, $\frac{\partial \Pi_D}{\partial v}=\frac{r \left(c^2 (1-\delta)^2+(1-r)^2 \left(v^2-1\right)\right)}{2 \beta  (1-r)^2 v^2}$. We can conclude that $\frac{\partial \Pi_D}{\partial v}<0$ if and only if $\underline{v}_1<v<\min\{\tilde{v}_1,\hat{v}\}$, and $\frac{\partial \Pi_D}{\partial v}>0$ if and only if $\tilde{v}_1<v<\hat{v}$. When $v>\hat{v}$, $\frac{\partial \Pi_D}{\partial v}=\frac{(2-\beta) r}{2}>0$. where $\tilde{v}_1=\frac{\sqrt{(1-r)^2-c^2 (1-\delta)^2}}{1-r}$. Therefore, the platform's profit function with respect to $v$ is first independent of $v$, then decreasing, and finally increasing.

When $0<v<\underline{v}_1$, $\frac{\partial \Pi_D}{\partial \beta}=0$, the profit is independent on $\beta$. When $\underline{v}_1<v<\hat{v}$, $\frac{\partial \Pi_D}{\partial \beta}=\frac{r \left(c^2 (1-\delta)^2-(1-r)^2 (v-1)^2\right)}{2 \beta ^2 (1-r)^2 v}$. We can conclude that $\frac{\partial \Pi_D}{\partial \beta}>0$ if and only if  $\underline{v}_1<v<\min\{\hat{v},\tilde{v}_2\}$, and $\frac{\partial \Pi_D}{\partial v}<0$ if and only if $\tilde{v}_2<v<\hat{v}$. 
where $\tilde{v}_2=\frac{1-r+c(1-\delta)}{1-r}$. When $v>\hat{v}$,  $\frac{\partial \Pi_D}{\partial \beta}=\frac{-r v}{2}<0$. Therefore, the platform's profit function with respect to $v$ is first independent of $v$, then increasing, and finally decreasing. 
$\blacksquare$

\subsection*{Proof of Proposition \ref{prop:2}}
The platform maximizes his profit under self-disclosure policy. According to the magnitude of $v$, we divide the analysis into three cases:

CASE 1: When $\underline{v}_1<v<\underline{v}_2$, According to Equation \ref{eqa:5}, no creator adopts $AINAI$ strategy, so the platform has no need to impose any penalty. 

CASE 2: When $\underline{v}_2<v<\hat{v}$, according to Equation \ref{eqa:6}:
\begin{enumerate}
\item $0<p<\overline{p}$, the platform's profit is a concave function with respect to $p$: 
\begin{align*}
\pi_D=
&-\frac{\beta r p^2}{2 k v (1-\beta  k)}+\frac{c (1-\delta) r p}{k (1-r) v}+\frac{(r-1)^2 v^2 \left(\beta  \left(\beta  \left(k^2-1\right)-k+2\right)-1\right)}{2 \beta  k (1-r)^2 v (\beta  k-1)}+\\&\frac{r \left(c^2 (1-\delta)^2 (1-\beta  k)+\beta  k (1-r)^2+2 (\beta -1) (1-r)^2 v (\beta  k-1)-(1-r)^2\right)}{2 \beta  k (r-1)^2 v (\beta  k-1)}
\end{align*}
We can reorganize $\frac{\partial \pi_D}{\partial p}\big|_{p=0}=\frac{c (1-\delta) r}{k (1-r) v}>0$, but $\frac{\partial \pi_D}{\partial p}\big|_{p=\overline{p}}=\frac{r (1-v-\beta  k+\beta  v)}{k v (1-\beta k)}$, the sign of this expression is ambiguous. when $\frac{1-\beta  k}{1-\beta}<v<\frac{1-r-c (1-\delta)}{(1-\beta) (1-r)}$, $\frac{\partial \pi_D}{\partial p}\big|_{p=\overline{p}}<0$, and $\pi_D$ increases in $p$ first and then decreases. Thus, the optimal penalty $p^*=\frac{c (1-\delta) (1-\beta  k)}{\beta  (1-r)}$, which is attained at $\frac{\partial \pi_D}{\partial p}=0$. When $\underline{v}_2<v\leq \min \{\frac{1-\beta  k}{1-\beta},\frac{1-r-c+c \delta}{1-r-\beta +\beta  r}\}$, $\frac{\partial \pi_D}{\partial p}\big|_{p=\overline{p}}>0$, and $\pi_D$ increases in $v$. Thus, the optimal penalty $p^*=\overline{p}$.

\item $p>\overline{p}$, $\Pi_D=\frac{r \left((1-r)^2 \left(v^2+1\right)-c^2 (1-\delta)^2\right)}{2 (1-r)^2 v}$, and the whole function is continuous.

\end{enumerate}

Therefore, we have the result of Case 2 is: 
\begin{equation}
\label{eqa:12}
p_1^*=
\begin{cases}
\frac{c (1-\delta ) (1-\beta  k)+(1-r) (\beta  k-1+(1-\beta)v)}{\beta  (1-r)} & \underline{v}_2<v\leq \min \{\tilde{v}_3,\hat{v}\}\\
\frac{c (1-\delta) (1-\beta  k)}{\beta  (1-r)}  & \tilde{v}_3<v<\hat{v}\\
\end{cases}
\end{equation}

CASE 3: When $\hat{v}<v<\overline{v}$ and, according to Equation \ref{eqa:7}:
\begin{enumerate}
\item When $0<p<\hat{p}$, the platform's profit is a concave function with respect to $p$: $\Pi_D=\frac{r \left(-\beta ^2 p^2+v^2 (2-\beta  (2-\beta +k))\right)}{2 v (1-\beta  k)}$, and $\frac{\partial \pi_D}{\partial p}\big|_{p=\hat{p}}=0$, so when $0<p<\hat{p}$, $\Pi_D$ decreases in $p$, and $p^*=0$.

\item When $\hat{p}<p<\overline{p}$, 
\begin{align*}
\pi_D=
&-\frac{\beta r p^2}{2 k v (1-\beta  k)}+\frac{c (1-\delta) r p}{k (1-r) v}+\frac{(r-1)^2 v^2 \left(\beta  \left(\beta  \left(k^2-1\right)-k+2\right)-1\right)}{2 \beta  k (1-r)^2 v (\beta  k-1)}+\\&\frac{r \left(c^2 (1-\delta)^2 (1-\beta  k)+\beta  k (1-r)^2+2 (\beta -1) (1-r)^2 v (\beta  k-1)-(1-r)^2\right)}{2 \beta  k (r-1)^2 v (\beta  k-1)}
\end{align*}
First, $\frac{\partial \pi_D}{\partial p}\big|_{p=\hat{p}}=\frac{r (1-r-v-\beta  c k+\beta  (c \delta  k+v)+(1-\beta) r v)}{k (1-r) v (1-\beta  k)}$, under the restriction on $\hat{v}<v<\overline{v}$, we can conclude that when $\hat{v}<v<\frac{1-r-\beta  c  k(1-\delta)}{(1-r)(1-\beta)}$, $\frac{\partial \pi_D}{\partial p}\big|_{p=\hat{p}}>0$, ad when $\frac{1-r-\beta  c  k(1-\delta)}{(1-r)(1-\beta)}<v<\overline{v}$, $\frac{\partial \pi_D}{\partial p}\big|_{p=\hat{p}}<0$. 

Second, $\frac{\partial \pi_D}{\partial p}\big|_{p=\overline{p}}=\frac{r (1-v-\beta  k+\beta  v)}{k v (1-\beta k)}$. We can recognize that under the restriction on $\hat{v}<v<\overline{v}$, $\frac{\partial \pi_D}{\partial p}\big|_{p=\overline{p}}>0$ if and only if $\hat{v}<v<\tilde{v}_3$ and $\frac{\partial \pi_D}{\partial p}\big|_{p=\overline{p}}<0$ if and only if $\min\{\hat{v},\tilde{v}_3\}<v<\overline{v}$. 

Thus, when $\hat{v}<v<\tilde{v}_3$, $\pi_D$ increases in $p$, and $p^*=\overline{p}$. When $\max\{\hat{v},\frac{1-k \beta}{1-\beta}\}<v<\frac{1-r-\beta  c  k(1-\delta)}{(1-r)(1-\beta)}$, the function $\pi_D$ increases and then decreases in $p$, and $p^*=\frac{c (1-\delta) (1-\beta  k)}{\beta  (1-r)}$, which is attained at $\frac{\partial \pi_D}{\partial p}=0$. When $\frac{1-r-\beta  c  k(1-\delta)}{(1-r)(1-\beta)}<v<\overline{v}$, $\pi_D$ decreases in $p$, $p^*=\hat{p}$.

\item When $p>\overline{p}$, $\Pi_D=\frac{r \left((1-r)^2 \left(v^2+1\right)-c^2 (1-\delta)^2\right)}{2 (1-r)^2 v}$, and the whole function is continuous. 
\end{enumerate}

Because when $0<p<\hat{p}$, $\Pi_D$ decreases in $p$, and $\Pi_D$ is continuous in the whole range. So $\hat{p}$ wouldn't be the optimal decision. So first, when $\frac{1-r-\beta  c  k(1-\delta)}{(1-r)(1-\beta)}<v<\overline{v}$, $p^*=0$. Second, when $\hat{v}<v<\tilde{v}_3$, we need to compare $\pi_D \big|_{p=0}=\frac{r v (2-\beta  (2-\beta +k))}{2-2 \beta  k}$ and $\pi_D \big|_{p=\overline{p}}=\frac{r \left((1-r)^2 \left(1+v^2\right)-c^2 (1-\delta)^2\right)}{2 (1-r)^2 v}$. We can conclude that when $\hat{v}<v<\min\{\tilde{v}_3,\tilde{v}_4\}$, $\Pi_D\big|_{p=\overline{p}}>\Pi_D\big|_{p=0}$. When $\tilde{v}_4<v<\tilde{v}_3$,  $\Pi_D\big|_{p=\overline{p}}<\Pi_D\big|_{p=0}$. Third, when $\max\{\hat{v},\tilde{v}_3\}<v<\frac{1-r-\beta  c  k(1-\delta)}{(1-r)(1-\beta)}$,  we need to compare $\pi_D \big|_{p=0}=\frac{r v (2-\beta  (2-\beta +k))}{2-2 \beta  k}$ and $\pi_D \big|_{p=\frac{c (1-\delta) (1-\beta k)}{\beta  (1-r)}}=\frac{r \left(\beta  c^2 (\delta -1)^2 k (\beta  k-1)+(1-r)^2 \left(v^2 \left(\beta  \left(\beta -\beta  k^2+k-2\right)+1\right)-\beta  k-2 (\beta -1) v (\beta  k-1)+1\right)\right)}{2 \beta  k (1-r)^2 v (1-\beta  k)}$. We can conclude that when $\max\{\hat{v},\tilde{v}_3\}<v<\tilde{v}_5$, $\Pi_D\big|_{p=\frac{c (1-\delta) (1-\beta k)}{\beta  (1-r)}}>\Pi_D\big|_{p=0}$ and when $\max\{\tilde{v}_5,\tilde{v}_3\}<v<\frac{1-r-\beta  c  k(1-\delta)}{(1-r)(1-\beta)}$, $\Pi_D\big|_{p=\frac{c (1-\delta) (1-\beta k)}{\beta  (1-r)}}>\Pi_D\big|_{p=0}$.

Where $\tilde{v}_3=\frac{1-k \beta}{1-\beta}$, $\tilde{v}_4=\frac{\sqrt{1-\beta  k} \sqrt{(1-r)^2-c^2 (1-\delta )^2}}{(1-\beta ) (1-r)}$, $\tilde{v}_5=\frac{1}{1-\beta }-\frac{c (1-\delta ) \sqrt{\beta  k}}{(1-\beta ) (1-r)}$.

Therefore, we have the result of Case 3 is: 
\begin{equation}
\label{eqa:13}
p_2^*=
\begin{cases}
\frac{c (1-\delta ) (1-\beta  k)+(1-r) (\beta  k-1+(1-\beta)v)}{\beta  (1-r)} & \text{if } \hat{v}<v<\min\{\tilde{v}_3,\tilde{v}_4\}\\
\frac{c (1-\delta) (1-\beta  k)}{\beta  (1-r)}  & \text{if } \max\{\hat{v},\tilde{v}_3\}<v<\tilde{v}_5\\
0 &\text{if} \tilde{v}_4<v<\tilde{v}_3||\max\{\tilde{v}_5,\tilde{v}_3\}<v<\overline{v}.
\end{cases}
\end{equation}
Let $\overline{p}=\frac{c (1-\delta ) (1-\beta  k)+(1-r) (\beta  k-1+(1-\beta)v)}{\beta  (1-r)}$ and $\tilde{p}=\frac{c (1-\delta) (1-\beta  k)}{\beta  (1-r)}$.

Combining the results in Equations \ref{eqa:12} and \ref{eqa:13}, we obtain the platform's optimal penalty decision as follows:
\begin{equation}
\label{eqa:14}
p^*= 
\begin{cases}
\frac{c (1-\delta ) (1-\beta  k)+(1-r) (\beta  k-1+(1-\beta)v)}{\beta  (1-r)} & \text{if } \underline{v}_2<v<\min\{\tilde{v}_3,\tilde{v}_4\}\\
\frac{c (1-\delta) (1-\beta  k)}{\beta  (1-r)}  & \text{if } \tilde{v_3}<v<\tilde{v}_5\\\
0 & \text{if } \tilde{v}_4<v<\tilde{v}_3||\max\{\tilde{v}_3,\tilde{v_5}\}<v<\overline{v}.\\
\end{cases}
\end{equation}
$\blacksquare$

\subsection*{Proof of Proposition \ref{prop:3}}
According to the platform's profit in Equation \ref{eqa:PiD}, when $\underline{v}_1<v<\min\{\tilde{v}_3,\tilde{v}_4\}$, $\frac{\partial \Pi_D}{\partial v}=\frac{r \left(c^2 (1-\delta)^2+(1-r)^2 \left(v^2-1\right)\right)}{2 (1-r)^2 v^2}$. The sign depends on $c^2 (1-\delta)^2+(1-r)^2 \left(v^2-1\right)$, which is a convex function. And we can conclude that $\frac{\partial \Pi_D}{\partial v}<0$ if and only if $\underline{v}_1<v<\frac{\sqrt{(1-r)^2-c^2 (1-\delta )^2}}{1-r}$ and  $\frac{\partial \Pi_D}{\partial v}>0$ if and only if $\frac{\sqrt{(1-r)^2-c^2 (1-\delta )^2}}{1-r}<v<\min\{\tilde{v}_3,\tilde{v}_4\}$.
$\frac{\partial \Pi_D}{\partial \beta}=0$.

When $\tilde{v}_3<v<\tilde{v}_5$, $\frac{\partial \Pi_D}{\partial v}=(1-r)^2 \left(1-\beta  \left(2-k-  \left(1-k^2\right)\beta\right)\right) v^2-r (1-\beta  k) \left(1-(2-r) r-\beta  c^2 (1-\delta)^2 k\right)$, which is a convex function and has a positive roots as $v_8=\frac{\sqrt{1-\beta  k} \sqrt{1+(r-2) r-c^2 \beta (1-\delta)^2 k}}{\sqrt{(1-r)^2 \left(1-\beta  \left(\beta  \left(k^2-1\right)-k+2\right)\right)}}$. $v_8>\tilde{v}_3$ always holds. Thus, $\frac{\partial \Pi_D}{\partial v}<0$ if and only if $\tilde{v}_3<v<\min\{v_8,\tilde{v}_5\}$ and $\frac{\partial \Pi_D}{\partial v}>0$ if and only if $v_8<v<\tilde{v}_5$.
$\frac{\partial \Pi_D}{\partial \beta}=\frac{r (\beta  k-\beta  v+v-1) (v (\beta  (2 k-1)-1)+1-\beta k)}{2 \beta ^2 k v (1-\beta  k)^2}<0$.

When $\tilde{v}_4<v<\tilde{v}_3||\max\{\tilde{v}_3,\tilde{v_5}\}<v<\overline{v}$, $\frac{\partial \Pi_D}{\partial v}=\frac{r (2-\beta  (2-\beta +k))}{2-2 \beta  k}>0$.
$\frac{\partial \Pi_D}{\partial \beta}=-\frac{(1-\beta) r v (2-(1+\beta) k)}{2 (1-\beta  k)^2}<0$.

Let $v_D=\frac{\sqrt{(1-r)^2-c^2 (1-\delta )^2}}{1-r}$.
$\blacksquare$

\subsection*{Proof of Proposition \ref{prop:4_restate}}

We compare the profit in the two different policies shown as Equations \ref{eqa:PiN} and \ref{eqa:PiD}:

(i)When $\underline{v}_1<v<\underline{v}_2$ and  $\underline{v}_2<v<\min\{\hat{v},\tilde{v}_3\}$, 
$\pi_N^*-\pi_D^*=\frac{(1-\beta) r ((1-r)^2 (v-1)^2-c^2 (1-\delta)^2)}{2 \beta  (1-r)^2 v}$, its sign depends on $(1-r)^2 (v-1)^2-c^2 (1-\delta)^2$, which is a convex function of $v$ with a root $v_0=\frac{1-r+c(1-\delta)}{1-r}$. For $\underline{v}_1<v<\underline{v}_2$, $\pi_N^*-\pi_D^*<0$ if and only if $\underline{v}_1<v<\min\{\underline{v}_2,v_0\}$,  and $\pi_N^*-\pi_D^*>0$ if and only if $v_0<v<\underline{v}_2$. For $\underline{v}_2<v<\min\{\hat{v},\tilde{v}_3\}$, $\pi_N^*-\pi_D^*<0$ if and only if $\underline{v}_2<v<\min\{\hat{v},\tilde{v}_3,v_0\}$, $\pi_N^*-\pi_D^*>0$ if and only if $\max\{\underline{v}_2,v_0\}<v<\min\{\hat{v},\tilde{v}_3\}$.

(ii) When $\tilde{v}_3<v<\hat{v}$, 
\begin{equation*}
\Delta \Pi=\frac{(1-\beta) (1-k) (1-(1+k))\beta)v^2}{k (1-\beta  k)}+\frac{2 (1-\beta) (1-k)}{k} v+1-\frac{1}{k}+\frac{(\beta -1) c^2 (\delta -1)^2}{(r-1)^2}
\end{equation*}
The two roots of the function are $v_1=\frac{(1-\beta  k) \left(k \left(\beta +\sqrt{\frac{(1-\beta) (1-k) \left((1-\beta) c^2 (1-\delta)^2 (\beta +\beta  k-1)+\beta ^2 (1-k) (1-r)^2\right)}{k (r-1)^2 (1-\beta  k)}}-1\right)+1-\beta\right)}{(1-\beta) (1-k) (1-(1+k)\beta)}$ and $v_2=\frac{(1-\beta  k) \left(k \left(\beta -\sqrt{\frac{(1-\beta) (1-k) \left((1-\beta) c^2 (1-\delta)^2 (\beta +\beta  k-1)+\beta ^2 (1-k) (1-r)^2\right)}{k (r-1)^2 (1-\beta  k)}}-1\right)+1-\beta\right)}{(1-\beta) (1-k) (1-(1+k)\beta)}$, If $1-(1+k))\beta>0$, $\Delta \Pi$ is a concave function of $v$, $v_1>v_2$, $v_1>\hat{v}$ always holds, so we can conclude that $\Delta \Pi<0$ if and only if $\tilde{v}_3<v<\min\{\hat{v},v_2\}$, and $\Delta \Pi>0$ if and only if $\max\{\tilde{v}_3,v_2\}<v<\hat{v}$. When $1-(1+k)\beta<0$, $v_1<v_2$, $v_1<\tilde{v}_3$ always holds, so we can conclude that $\Delta \Pi<0$ if and only if $\tilde{v}_3<v<\min\{\hat{v},v_2\}$ and $\Delta \Pi>0$ if and only if $\max\{\tilde{v}_3,v_2\}<v<\hat{v}$. 

(iii) when $\hat{v}<v<\min \{\tilde{v}_3\,\tilde{v}_4\}$, 
\begin{equation*}
\Delta \Pi=\frac{r \left(c^2 (1-\delta)^2-(1-r)^2 \left(1-(1-\beta ) v^2\right)\right)}{2 (1-r)^2 v}
\end{equation*}

The sign of $\Delta \Pi$ depends on $c^2 (1-\delta)^2-(1-r)^2 (1-(1-\beta ) v^2)$,  which is a convex function with the only position root $v_5=\frac{\sqrt{(1-r)^2-c^2 (1-\delta)^2}}{\sqrt{(1-\beta) (1-r)^2}}$. Further, we have $v_5<\tilde{v}_4$ always holds.

We can conclude that  $\Delta \Pi<0$ if and only if $\hat{v}<v<\min\{v_5,\tilde{v}_3\}$, and $\Delta \Pi>0$ if and only if $\min\{v_5,\hat{v}\}<v<\min\{\tilde{v}_3,\tilde{v}_4\}$.

(iv) when $\max\{\hat{v},\tilde{v}_3\}<v<\tilde{v}_5$, 
\begin{equation*}
\Delta \Pi=\frac{1}{2} (2-\beta) r v-\frac{r \left(\beta  c^2 (\delta -1)^2 k (\beta  k-1)+(1-r)^2 \left(v^2 \left(\beta  \left(\beta -\beta  k^2+k-2\right)+1\right)-\beta  k-2 (\beta -1) v (\beta  k-1)+1\right)\right)}{2 \beta  k (1-r)^2 v (1-\beta  k)}
\end{equation*}
The sign of this function is same as $f(v)=(1-\beta) (1-r)^2 r (\beta  (1+k (1-\beta  k))-1) v^2+2 (1-\beta) (1-r)^2 (1-\beta  k)rv-\beta  c^2 (\delta -1)^2 k r (\beta  k-1)+\beta  k r (r-1)^2-(r-1)^2 r$. It has two roots as:\\$v_3=\frac{\beta ^2 (-k) (r-1)^2+\beta  (k+1) (r-1)^2-(r-1)^2+\sqrt{(\beta -1) \beta  k (r-1)^2 (\beta  k-1) \left(c^2 (\delta -1)^2 (\beta  (k (\beta  k-1)-1)+1)-\beta  (k-1) (r-1)^2\right)}}{(\beta -1) (r-1)^2 (\beta  (k (\beta  k-1)-1)+1)}$ and \\$v_4=\frac{\beta ^2 (-k) (r-1)^2+\beta  (k+1) (r-1)^2-(r-1)^2-\sqrt{(\beta -1) \beta  k (r-1)^2 (\beta  k-1) \left(c^2 (\delta -1)^2 (\beta  (k (\beta  k-1)-1)+1)-\beta  (k-1) (r-1)^2\right)}}{(\beta -1) (r-1)^2 (\beta  (k (\beta  k-1)-1)+1)}$. 

When $\beta  (1+k (1-\beta  k))-1>0$, $v_3>v_4$ it's a convex function of $v$ and $v_4<\min\{\hat{v},\tilde{v}_3\}$ and $v_3<\tilde{v}_5$ always holds. So $\Delta \Pi<0$ if and only if $\max\{\hat{v},\tilde{v}_3\}<v<v_3$, and $\Delta \Pi>0$ if and only if $\max\{\hat{v},\tilde{v}_3,v_3\}<v<\tilde{v}_5$. 
When $\beta  (1+k (1-\beta  k))-1<0$, $v_3<v_4$, it's a concave function of $v$ and $v_3<\tilde{v}_5$ and $v_4>\tilde{v}_5$ always holds. So $\Delta \Pi<0$ if and only if $\max\{\hat{v},\tilde{v}_3\}<v<v_3$, and $\Delta \Pi>0$ if only if $\max\{\hat{v},\tilde{v}_3,v_3\}<v<\tilde{v}_5$.

(v) when $\tilde{v}_4<v<\tilde{v}_3||\max\{\tilde{v}_3,\tilde{v}_5\}<v<\overline{v}$, $\pi_N^*-\pi_D^*=\frac{\beta  r v (1-k)(1-\beta)}{2-2 \beta  k}>0$. 

Further, we analyze the boundaries $v=\{v_0,v_2,v_3,v_5\}$. It is easy to see that $\frac{\partial v_5}{\partial \delta}>0$, $\frac{\partial v_3}{\partial \delta}>0$, $\frac{\partial v_2}{\partial \delta}<0$ and $\frac{\partial v_0}{\partial \delta}<0$. The root of $v_3=v_5$ is $\delta=1-\frac{(1-r) \sqrt{(1-\beta)  (k (2-\beta  k)-1)\beta}}{(1-\beta ) c}$, the root of $v_2=v_0$ is $\delta=1-\frac{(1-k) (1-r)\beta }{(1-\beta) c}$.

So given the continuity of the boundaries, we obtain:
only when $\max\{\delta_1,0\}<\delta<\min\{\delta_0,\delta_2\}$, disclosure policy realize higher profit. Where $\delta_0$ is the root of $v=\underline{v}_1$. When $0<\delta<1-\frac{(1-r) \sqrt{(1-\beta)  (k (2-\beta  k)-1)\beta}}{(1-\beta ) c}$), $\delta_1$ is the root of $v=v_5$. When $1-\frac{(1-r) \sqrt{(1-\beta)  (k (2-\beta  k)-1)\beta}}{(1-\beta ) c}<\delta<1-\frac{(1-k) (1-r)\beta }{(1-\beta) c}$, $\delta_1$ is the root of $v=v_3$ and $\delta_2$ is the root of $v=v_2$. When $1-\frac{(1-k) (1-r)\beta }{(1-\beta) c}<\delta<1$, $\delta_2$ is the root of $v=v_0$. 
$\blacksquare$

\subsection*{Proof of Proposition \ref{disclosureoncreator}}
According to the Equation \ref{eqa:CS_restate}, we can derive the creator's surplus in the two cases are:
\begin{equation*}
\label{eqa:CSN}
CS_N^*= 
\begin{cases}
\frac{c^2 (\delta -1)^2+2 c (r-1) (-\delta +v (\beta +\delta -1)+1)+(r-1)^2 \left(2 \beta  v+(v-1)^2\right)}{2 \beta  (1-r) v} & \underline{v_1}<v<\hat{v}\\
\frac{1}{2} (2-\beta) (1-r) v-c \delta& \hat{v}<v<\overline{v}.\\
\end{cases}
\end{equation*}

\begin{equation*}
\label{eqa:CSD}
CS_D^*= 
\begin{cases}
\frac{c^2 (\delta -1)^2+2 c (r-1) (\delta  (v-1)+1)+(r-1)^2 \left(v^2+1\right)}{2 (1-r) v} & \underline{v_1}<v<\min\{\tilde{v}_3,\tilde{v}_4\}\\
\frac{2 r \left((\beta  k-1) (\beta  c (\delta -1) k+1)-v (\beta  k-1) (-2 \beta +\beta  c \delta  k+2)+v^2 \left(\beta  \left(\beta  \left(k^2-1\right)-k+2\right)-1\right)\right)}{2 \beta  k (r-1) v (\beta  k-1)}\\-\frac{(\beta  k-1) (\beta  c (\delta -1) k (c (\delta -1)+2)+1)+2 v (\beta  k-1) (-\beta +\beta  c \delta  k+1)}{2 \beta  k (r-1) v (\beta  k-1)}\\+\frac{r^2 \left(v^2 \left(\beta  \left(\beta -\beta  k^2+k-2\right)+1\right)-\beta  k-2 (\beta -1) v (\beta  k-1)+1\right)+v^2 \left(\beta  \left(\beta -\beta  k^2+k-2\right)+1\right)}{2 \beta  k (r-1) v (\beta  k-1)}  & \tilde{v_3}<v<\tilde{v}_5\\
\frac{(1-r) v \left(\beta ^2-\beta  (k+2)+2\right)}{2-2 \beta  k}& \tilde{v}_4<v<\tilde{v}_3||\max\{\tilde{v}_3,\tilde{v_5}\}<v<\overline{v}.\\
\end{cases}
\end{equation*}

We compare the creator's surplus in the two cases $\Delta CS=CS_N^*-CS_D^*$:

(i) When $\underline{v}_1<v<\underline{v}_2$ and $\underline{v}_2<v<\min\{\hat{v},\tilde{v}_3\}$, $\Delta CS=\frac{(1-\beta) (1+\delta c-c+r v-r-v)^2}{2 \beta  (1-r) v}>0$

(ii) When $\tilde{v}_3<v<\hat{v}$, $\Delta CS=\frac{(k-1) (r-1)^2 (\beta  k-(\beta -1) v (-2 \beta  k+v (\beta +\beta  k-1)+2)-1)}{2 \beta  k (1-r) v (\beta  k-1)}-\frac{2 (\beta -1) c (\delta -1) (r-1) (v-1)-(\beta -1) c^2 (\delta -1)^2}{2 \beta  (1-r) v}>0$.

(iii) When $\hat{v}<v<\min \{\tilde{v}_3\,\tilde{v}_4\}$, $\Delta CS=\frac{2 c (1-r)(1-\delta)+(1-r)^2 ((1-\beta) v^2-1)-c^2 (1-\delta)^2}{2 (1-r) v}>0$.

(iv) When $\max\{\hat{v},\tilde{v}_3\}<v<\tilde{v}_5$, 
\begin{align*}
\Delta CS=&\frac{\beta  c (\delta -1) k (c (\delta -1)+2)-2 r (\beta  c (\delta -1) k+1)+r^2+1}{2 \beta  k (r-1) v}+\frac{2 (1-\beta ) (1-r)^2}{2 \beta  k (1-r)}\\&+\frac{(\beta -1) (r-1)^2 v (\beta  (k (\beta  k-1)-1)+1)}{2 \beta  k (1-r) (1-\beta  k)}>0
\end{align*}

(v) when $\tilde{v}_4<v<\tilde{v}_3||\max\{\tilde{v}_3,\tilde{v}_5\}v<\overline{v}$, $\Delta CS=\frac{(1-\beta) \beta  (1-k) (1-r) v}{2-2 \beta  k}>0$.

Therefore, $CS_N \geq CS_D^*$ always holds.
$\blacksquare$

\subsection*{Proof of Proposition \ref{qualitytransparency}}
According to the Equation \ref{eqa:T_restate}, we can derive the creator's surplus in the two cases are:
\begin{equation*}
\label{eqa:CSN}
T_N^*= 
\begin{cases}
\frac{(1-r) (1-\beta +(2 \beta -1) v)-(1-\beta) (1-\delta) c}{\beta  (1-r) v} & \underline{v}_1<v<\hat{v}\\
\beta & \hat{v}<v<\overline{v}.\\
\end{cases}
\end{equation*}

\begin{equation*}
\label{eqa:TD}
T_D^*= 
\begin{cases}
\frac{1-v-\beta ^2 (k (k v-1)+v)+\beta  ((k+2) v-k-1)}{\beta  k v (1-\beta  k)} & \tilde{v_3}<v<\tilde{v}_5\\
\frac{\beta  (2-\beta -k)}{2-\beta  k}& \tilde{v}_4<v<\tilde{v}_3||\max\{\tilde{v}_3,\tilde{v_5}\}<v<\overline{v}.\\
\end{cases}
\end{equation*}

We compare the platform's content transparency in the two cases $\Delta T=T_N^*-T_D^*$

(i) When $\underline{v}_1<v<\underline{v}_2$ and $\underline{v}_2<v<\min\{\hat{v},\tilde{v}_3\}$, $\Delta T=\frac{(1-\beta) ((1-r) (1-v)-c (1-\delta))}{\beta  (1-r) v}<0$

(ii) When $\tilde{v}_3<v<\hat{v}$, 
\begin{align*}
\Delta T=&\frac{(\beta -1) (c (\delta -1)+r-1)}{\beta  k (r-1) v}+\frac{v \left(\beta  \left(-\beta  \left(k^2+1\right)+k+2\right)-1\right)}{\beta  k v (\beta  k-1)}\\&+\frac{-\beta +(2 \beta -1) v+1}{\beta  v}<0
\end{align*}

(iii) When $\hat{v}<v<\min \{\tilde{v}_3\,\tilde{v}_4\}$, $\Delta T=\beta-1<0$.

(iv) When $\max\{\hat{v},\tilde{v}_3\}<v<\tilde{v}_5$, 
\begin{equation*}
\Delta T=\frac{(1-\beta) \left(\beta ^2 k^2 v+\beta  k-\beta  (k+1) v+v-1\right)}{\beta  k v (1-\beta  k)}<0
\end{equation*}

(v) when $\tilde{v}_4<v<\tilde{v}_3||\max\{\tilde{v}_3,\tilde{v}_5\}<v<\overline{v}$, $\Delta T=-\frac{(1-\beta) (1-k) \beta }{1-\beta  k}<0$.

Therefore, $T_N^*\leq T_D^*$ always holds.

According to the Equation \ref{eqa:Q_restate}, we can derive the creator's surplus in the two cases are:
\begin{equation*}
\label{eqa:CSN}
Q_N^*= 
\begin{cases}
\frac{c (\delta -1) (v-1)+(r-1) (v (\beta +v-2)+1)}{\beta  (r-1) v} & \underline{v_1}<v<\hat{v}\\
v & \hat{v}<v<\overline{v}.\\
\end{cases}
\end{equation*}

\begin{equation*}
\label{eqa:CSD}
Q_D^*= 
\begin{cases}
\frac{c (\delta -1) (v-1)}{(r-1) v}+v+\frac{1}{v}-1 & \underline{v}_1<v<\min\{\tilde{v}_3,\tilde{v}_4\}\\
\frac{\beta  c k+\beta  c \delta  k (v-1)-v (\beta +\beta  c k+\beta  (k-1) v+v-2)+r v (\beta +\beta  (k-1) v+v-2)+r-1}{\beta  k (r-1) v} &  \tilde{v}_3<v<\tilde{v}_5\\
v& \tilde{v}_4<v<\tilde{v}_3||\max\{\tilde{v}_3,\tilde{v}_5\}<v<\overline{v}.\\
\end{cases}
\end{equation*}

We compare the content quality in the two cases $\Delta Q=Q_N^*-Q_D^*$

(i) When $\underline{v}_1<v<\min\{\hat{v},\tilde{v}_3\}$, $\Delta Q=\frac{(\beta -1) (v-1) (c (\delta -1)+(r-1) (v-1))}{\beta  (1-r) v}$, which is a convex function of $v$ and the two roots are $v_6=\frac{1-r-c(1-\delta)}{1-r}$ and $v_7=1$. $v_6<\underline{v}_1$. Thus, we can derive that $\Delta Q<0$ if and only if $\underline{v}_1<v<\min\{\hat{v},1\}$ and $\Delta Q>0$ if and only if $1<v<\min\{\hat{v},\tilde{v}_3\}$.

(ii) When $\tilde{v}_3<v<\hat{v}$, 
\begin{align*}
\Delta Q=\frac{(v-1) (k ((\beta -1) c (\delta -1)+(r-1) ((\beta -1) v+1))-(r-1) ((\beta -1) v+1))}{\beta  k (1-r) v}>0.
\end{align*}

(iii) When $\hat{v}<v<\min \{\tilde{v}_3,\tilde{v}_4\}$, $\Delta Q=\frac{(v-1) (1-r-c(1-\delta)}{(1-r) v}$. So we can have $\Delta Q>0$ if and only if $1<v<\min \{\tilde{v}_3,\tilde{v}_4\}$ and $\Delta Q<0$ if and only if $\hat{v}<v<\min\{\tilde{v}_4,1\}$.

(iv) When $\max\{\hat{v},\tilde{v}_3\}<v<\tilde{v}_5$, 
\begin{equation*}
\Delta Q=\frac{(v-1) (1-r-v-\beta  c k+\beta  (c \delta  k+v)+(1-\beta) r v)}{\beta  k (1-r) v}>0.
\end{equation*}

(v) when $\tilde{v}_4<v<\tilde{v}_3||\max\{\tilde{v}_3,\tilde{v}_5\}<v<\overline{v}$, $\Delta Q=0$. 

Therefore, $Q_N^*<Q_D^*$ if and only if $\underline{v}_1<\min\{\tilde{v}_4,1\}$ and $Q_N^*>Q_D^*$ if and only if $1<v<\tilde{v}_4||1<v<\tilde{v}_5$. Otherwise, $\Delta Q=0$
$\blacksquare$

\end{document}